\documentclass{appolb}
\usepackage{epsfig}

\begin{document}
\title{Decay widths of charmonia in a hot equilibrated medium%
\thanks{Presented at Three Days of Strong Interactions, Wroclaw (Poland)    9. - 11. VII. 2009}%
}
\author{Massimo Mannarelli
\address{Instituto de Ciencias del Espacio (IEEC/CSIC) 
Campus Universitat Aut\` onoma de Barcelona,
Facultat de Ci\` encies, Torre C5 
E-08193 Bellaterra (Barcelona), Spain}
\and
Floriana Giannuzzi
\address{I.N.F.N., Sezione di Bari, I-70126 Bari, Italia, 
Universit\`a degli Studi di Bari, I-70126 Bari, Italy}
}
\maketitle
\begin{abstract}
We investigate the properties of charmonia in a thermal medium, showing that with
increasing temperature the decay widths of these mesons behave in a non-trivial way.
Employing a potential model with interaction potential extracted from thermal lattice
QCD calculations of the free-energy of a static quark-antiquark pair, we study some
decay processes in the crossover region. We find that at temperatures $T \sim T_c$ the
decay widths of the $J/\Psi$ that depend on the value of the wave function at the origin
are enhanced with respect to the values in vacuum. In the same temperature range the decay width of the process $\chi_{cJ} \to J/\Psi + \gamma$ is enhanced by approximately a factor
$6$ with respect to the value in vacuum. At higher temperatures the charmonia states
dissociate and the widths of both decay processes become vanishing small.
\end{abstract}
  
\section{Introduction}  
The heavy-ion collision program aims  to identify and characterize the properties of deconfined quark matter at high temperatures. 
Many valuable probes of the properties of  the  medium produced in a heavy-ion collision  are available, which include   jets, electromagnetic signals and heavy  quarkonia states ($Q \bar Q$)~\cite{Abreu:2007kv}. In particular much work has been devoted to understand how the presence  of deconfined quarks and gluons may affect the binding  of quarkonia~\cite{reviews,Hatsuda:1985eb,Matsui:1986dk}. 

The analyses of the binding energies of mesonic states as a function of the temperature have been carried out by various authors using potential models, with  interaction potentials extracted from lattice QCD calculations~\cite{Shuryak:2004tx,Wong:2004zr,Mocsy:2007jz,Mannarelli:2005pz}. These analyses show that  quarkonia dissociate at  temperatures close to the critical temperature of QCD, $T_c$.  Analogous results have been obtained by studying the correlation functions of charmonia above deconfinement~\cite{Asakawa:2003re}.
 
In the present study~\cite{Giannuzzi:2009gb} we are interested in determining how the decay widths for the leptonic, hadronic and radiative channels are influenced by the presence of the thermal medium. In order to  determine the  decay widths  we first evaluate the radial wave function of the pertinent charmonia states   employing the non-relativistic Scrh\"odinger equation (and for comparison  the relativistic Salpeter equation)   with interaction   potential extracted from thermal lattice QCD simulations.  

The remarkable result of our analysis is that some decay widths change drastically close to $T_c$.
In the transition region  we find that  the decay widths of the $J/\Psi$  that depend on the value of the wave function at the origin change by approximately a factor $2$ with respect to the corresponding values in  vacuum. We also investigate the properties of the $\chi_c$ meson and in particular of the  $\chi_{c J} \to J/\Psi + \gamma$ transition. This  radiative decay contributes to  the total inclusive $J/\Psi$ production by a fraction of about  $0.3$, as determined in proton proton and $\pi^+$ $\pi^-$  data~\cite{Antoniazzi:1992iv}.   We find that this $J/\Psi$ production mechanism is enhanced close to the critical temperature by approximately a factor $6$  with respect to the value in  vacuum.


\section{The method}
Potential models have been quite successful in describing the properties of heavy quark bound states in vacuum (see {\it e.g.} \cite{Appelquist:1978aq}). Our key assumption is that at any temperature $T$, the interaction between heavy quarks can be  approximated by an instantaneous potential, $V(r, T)$, where $r$ is the radial coordinate.

We shall study bound states of heavy quarks by using  the non relativistic    Schr\"odinger equation 
\begin{equation}\label{schrodinger}
\left( 2M_Q - \frac{\nabla^2}{M_Q} + V(r, T) \right) \psi_i = E_i  \psi_i  \,,
\end{equation}
where $M_Q$ is the constituent  mass of the heavy quark; $\psi_i$ and $ E_i$ represent the wave function and the mass of the corresponding $Q\bar Q$ state, respectively.
Comparing the results of this analysis with those obtained using the   Salpeter  equation for the $S-$wave states 
\begin{equation}\label{salpeter}
\left(2\sqrt{M_Q^2 - \nabla^2}  + V(r,T) \right)\psi_i = M_i\psi_i \,,
\end{equation}
we  find excellent agreement between the outcomes of the two methods, meaning that for these mesonic states relativistic corrections are small.

There are some controversies about how to extract the potential $V(r,T)$   from the color-singlet free-energy $F_1(r,T)$  (see  \cite{Wong:2004zr,Mocsy:2007jz,Satz:2008zc}). In principle, the internal energy $U_1$ is obtained 
subtracting the entropy contribution from the free-energy
\begin{equation}\label{potU}
U_1\,=\,F_1-T\frac{{\rm \partial} F_1}{{\rm \partial} T} \,.
\label{U1}
\end{equation}

However, this subtraction  procedure has been put to question, one of the reasons being that  around the critical temperature,  the potential $U_1$  is more attractive than the potential at zero temperature. One possible interpretation of this result is that
at non-vanishing temperatures there can be  additional interactions between the two static quarks~\cite{Satz:2008zc} .  Below $T_c$ this effect should be due to hadrons and could be related to the  string ``flip-flop" interaction~\cite{Miyazawa:1979vx}; above $T_c$ it should be related to the antiscreening properties of QCD. 
In the following and in agreement with Ref.~\cite{Satz:2008zc} we shall use the internal energy in  Eq.~(\ref{U1}) as the interaction potential and show the results obtained using the free-energy only for comparison. As we shall see, using $U_1$, we obtain dissociation temperatures for charmonia consistent with lattice results obtained with the Maximum Entropy Method~\cite{Asakawa:2003re}. Employing the free-energy as interaction potential gives much smaller dissociation temperatures. 


We shall consider the free-energy  proposed in~\cite{Dixit:1989vq} employing the formalism of the Debye-H\"uckel theory, with the parameterization of Ref.~\cite{Digal:2005ht}, 
\begin{equation}\label{Fdixit}
F_1(r,T) = \frac{\sigma}{\mu} \left[ \frac{\Gamma(1/4)}{2^{3/2}\Gamma(3/4)} - \frac{\sqrt{x}}{2^{3/4}\Gamma(3/4)}K_{1/4}(x^2 + \kappa\, x^4)\right]-\frac{\alpha}{r}(\exp(-x)+x) \,, 
\end{equation}
where $K_{1/4}$ is the modified Bessel function, $x=\mu \,r$, while the functions  $\mu \equiv \mu(T)$ and $\kappa \equiv \kappa(T)$ are determined by fitting the lattice data.  Once these two functions are fixed, this parameterization  of the free-energy is in excellent agreement with lattice data for $T \le 2\, T_c$. 
At short distances the static quark-antiquark  free-energy is normalized in such a way that it reproduces the free-energy at zero temperature
\begin{equation} \label{cornell}
F(r, T = 0) =-\frac{\alpha}r +\sigma r \,,
\end{equation}
where $\sigma $ is the string tension and $\alpha$ is the Coulomb coupling constant. The values of these two parameters can be phenomenologically fixed; we shall take   $\sigma = 0.16 $ GeV$^2$, $\alpha=0.2$ and with $m_c=1.28$ GeV one obtains $M_{J/\psi} \simeq 3.11$ GeV and  $M_{\chi} \simeq 3.43 $ GeV.



\section{ Masses and Decay widths}
In order to determine the decay widths it is necessary to compute the wave function and the mass of the charmonia states. For $l=0$ states, they are determined solving both the Schr\"odinger equation and the Salpeter equation through the Multhopp method  while for $P-$wave states we only solve the Schr\"odinger equation. 

We have considered various decay processes. Annihilation  processes of charmonia can be viewed as two-stage 
factorized processes.  The first process consists in  $c \bar c$  annihilation into gauge bosons. The annihilation takes place at the characteristic distance $r$ of order $1/m_c$, so for a non-relativistic pair $r \to 0$ and the annihilation amplitude is proportional to the wave function at the origin. Then, the produced  gauge boson  decays   or fragments into leptons and  hadrons.  In vacuum it is assumed that the  inclusive probability of the latter process is equal to one. At finite temperature we  assume that such a factorization persists. Therefore, the width of   hadronic  and  leptonic decays will depend on temperature  exclusively through the radial wave function at the origin and  the mass of the meson~\cite{Appelquist:1978aq} and can be expressed as
\begin{equation}\label{width1}
\Gamma_A \propto \Gamma_{\ell^+\ell^-}  \propto \Gamma(^3\!S_1 \to 3 g) \propto    \frac{|R_{s}(T,0)|^2}{M(T)^2}\,.
\end{equation}
We also consider the radiative transitions $\chi_{c J} \to J/\Psi + \gamma$ from the $^3 P_J$ levels to the $^3S_1$ state with rate 
\begin{equation}\label{width2}
\Gamma_B \propto \Gamma_{\chi_{cJ} \to J/\Psi + \gamma}  \propto (2 J +1)  |I_{\rm PS}|^2\,,
\end{equation}
where $ I_{\rm PS}$ is the  overlap integral between the radial wave functions of the corresponding states.

\subsection{Decay widths at different temperatures}
Now we show the results  for dissociation temperatures and  decay widths obtained  using as $Q\bar Q$ potential the  internal energy  and for comparison the  free-energy.

Using the free-energy as interaction potential,  the $J/\psi$ dissociates at temperatures close to $1.2 \, T_c$, while   the $\chi_c$  dissociate at about  $ 0.95 \, T_c$. Using the internal energy as potential we find that the $J/\psi$ dissociates at approximately $1.8 \, T_c$, while the  $\chi_c$ dissociates at about $1.15 \, T_c$.   Using the internal energy   one obtains higher dissociation temperatures in agreement with the fact that $ U_1$ is more attractive than $F_1$. 

\begin{figure}[thdp]
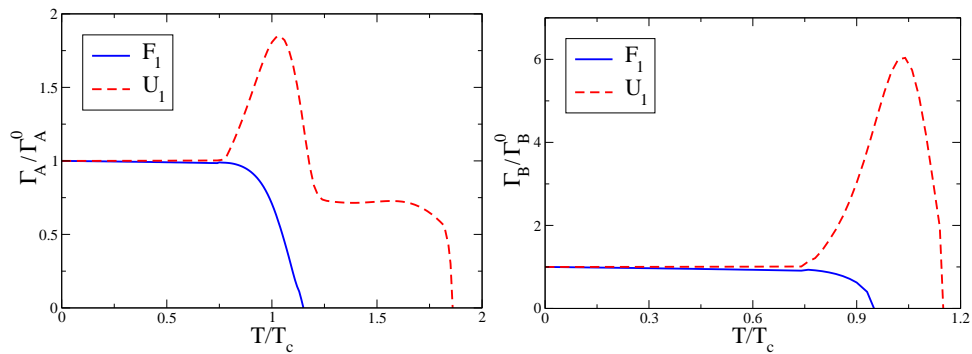

\includegraphics[width=2.5in,angle=-0]{Widthpsi.eps}
\includegraphics[width=2.5in,angle=-0]{Widthchi.eps}
\caption{(color online) Charmonia decay widths given in Eq.~(\ref{width1}) (left panel) and in Eq.~(\ref{width2}) (right panel). We have used  $F_{1}$, full blue lines, and $ U_{1}$, dashed red lines,  as potentials with the paramterization of  Eq.~(\ref{Fdixit}), and normalized the widths to the vacuum value.} \label{jpsi-origin}
\end{figure}

In Fig.~\ref{jpsi-origin} we report  the values of the decay width for the process in Eq.~(\ref{width1}), left panel,  and for the process in Eq.~(\ref{width2}), right panel,  as a function of the temperature. 
Employing $F_1$ one finds a monotonic decrease of the width that can be viewed  as due to a decrease of the screening length of the potential.
On the other hand,  when using  $ U_1$ we find that  across the transition region both decay widths become large. For the decays in Eq.~(\ref{width1}) the width is about a factor $2$ larger than in  vacuum. For the radiative decay in Eq.~(\ref{width2}), the ratio between the width in the thermal medium and in vacuum reaches a factor $6$ across the transition region. This effect is not due to the screening of the potential, but is rather connected with the ``cloud-cloud" interaction~\cite{Satz:2008zc}.  

In order to have a rough estimate of the effect, let us suppose that the $\chi_c$ is produced in the early stage of the heavy-ion collision and then travels for about $4$ fm in the thermally equilibrated medium. If the temperature is sufficiently low, so that  the decay width is approximately the same as in vacuum,  one has that less than $1 \%$ of   $\chi_c$ decay. On the other hand, for temperatures close to  $ T_c$, one should observe that about $5\%$  of $\chi_c$ decay. Since for temperatures close to $T_c$ one can neglect interactions of charmonia  with in-medium hadrons and gluons~\cite{reviews} the radiative decay process  should give a sizable contribution to the total decay width of the $\chi_c$. As regard the inclusive width of the  $J/\Psi$, one should consider that  at temperatures below $T_c$,  the process in Eq.~(\ref{width1}) gives approximately the inclusive width. However, at larger temperatures the dominant contribution  should be  due to the interaction with in-medium partons~\cite{reviews}.

The largest uncertainty in our calculations resides
in the extraction of the potential from   
lattice QCD calculations and pertinent parametrizations to numerically evaluate
the  entropy.  In order to test the robustness of our results one can employ a different  parametrizations of the internal energy. Using the expression of  $U_1(r,T)$  reported in Ref.~\cite{Wong:2004zr}   we find approximately the same results reported here. 

The analysis reported here can easily be extended to bottomia, moreover it  would be of some  interest to include   $D$-mesons.\\

\noindent\underline{\it Acknowledgment}\ This work was supported by the Spanish grant
FPA2007-60275, by the contract FPA2008-03918-E/INFN and  by the EU Contract No. MRTN-CT-2006-035482, "FLAVIAnet".


\end{document}